\documentclass[graybox]{svmult}

\usepackage{mathptmx}       % selects Times Roman as basic font
\usepackage{helvet}         % selects Helvetica as sans-serif font
\usepackage{courier}        % selects Courier as typewriter font
\usepackage{type1cm}        % activate if the above 3 fonts are
\usepackage{makeidx}         % allows index generation
\usepackage{graphicx}        % standard LaTeX graphics tool
\usepackage{multicol}        % used for the two-column index
\usepackage[bottom]{footmisc}% places footnotes at page bottom

\usepackage{braket} % by authors
\usepackage{amsfonts}

\makeindex             % used for the subject index
\begin{document}

\title*{Optical Hybrid Quantum Information Processing}
\author{Shuntaro Takeda and Akira Furusawa}
\institute{Shuntaro Takeda \at Department of Applied Physics, School of Engineering, The University of Tokyo, 7-3-1 Hongo, Bunkyo-ku, Tokyo 113-8656, Japan, \email{takeda@alice.t.u-tokyo.ac.jp}
\and Akira Furusawa \at Department of Applied Physics, School of Engineering, The University of Tokyo, 7-3-1 Hongo, Bunkyo-ku, Tokyo 113-8656, Japan,  \email{akiraf@ap.t.u-tokyo.ac.jp}}

\maketitle

\abstract*{Historically, two complementary approaches to optical quantum information processing have been pursued: 
qubits and continuous-variables, each exploiting either particle or wave nature of light.
However, both approaches have pros and cons.
In recent years, there has been a significant progress in combining both approaches
with a view to realizing hybrid protocols that overcome the current limitations.
In this chapter, we first review the development of the two approaches with a special focus on quantum teleportation and its applications.
We then introduce our recent research progress in realizing quantum teleportation by a hybrid scheme,
and mention its future applications to universal and fault-tolerant quantum information processing.}

\abstract{Historically, two complementary approaches to optical quantum information processing have been pursued: 
qubits and continuous-variables, each exploiting either particle or wave nature of light.
However, both approaches have pros and cons.
In recent years, there has been a significant progress in combining both approaches
with a view to realizing hybrid protocols that overcome the current limitations.
In this chapter, we first review the development of the two approaches with a special focus on quantum teleportation and its applications.
We then introduce our recent research progress in realizing quantum teleportation by a hybrid scheme,
and mention its future applications to universal and fault-tolerant quantum information processing.}

%%% Section 1 %%%%%%%%%%%%%%%%%%%%%%%%%%%%%%%%%%%%%%%%%%%%%%%%%%%%%%%%%%%%%%%%%%%%%%%%%%%%%%%%%%%%%%

\section{Introduction}
\label{sec:introduction}

\begin{figure}[t]
% \sidecaption
\centering
\includegraphics[scale=0.35]{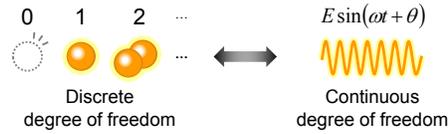}
\caption{Discrete and continuous degrees of freedoms of light.}
\label{fig:CV_DV}
\end{figure}

Optical quantum systems are one of the most promising candidates for quantum information processing (QIP)
since their decoherence is almost negligible under ambient conditions at room temperatures. 
This advantage, together with mature optical technologies such as beam splitters and nonlinear optical crystals,
enabled significant progress in the field of optical quantum communication and quantum computing. 
This progress was made by two complementary approaches, each exploiting only one aspect of the wave-particle duality of light (Fig.~\ref{fig:CV_DV}).
One utilizes the particle-like discrete nature of light to encode quantum information based on quantum bits (qubits)~\cite{07Kok,12Pan}. 
The other, which harnesses wave-like continuous nature of light, is based on continuous variables (CVs)~\cite{05Braunstein}.
The conceptual difference between these two approaches is analogous to classical digital (discrete) and analog (continuous) signal processing.

%%%%
Both approaches have advantages and disadvantages in terms of the practical realization of optical QIP. 
Qubit QIP enables high fidelity of operations,
but experimental realizations have been probabilistic and mostly required post-selection of successful events.
This is due to the low creation and detection efficiencies of photonic qubits.
On the other hand, CV QIP is deterministic, thanks to on-demand entanglement resources and efficient homodyne measurement; however, the fidelity of operations is limited by the imperfection of the entanglement.
In recent years, there has been significant progress in combining both technologies
with a view to realizing hybrid protocols that overcome the current limitations of optical QIP~\cite{11vanLoock,11Furusawa}.
This hybrid approach is analogous to the digital and analog hybrid signal processing in classical information processing: 
we set thresholds to digitize the originally continuous voltage signals. 
In optical QIP, this threshold was given by nature as the wave-particle duality of light. 
Therefore it naturally follows to take advantage of both features in optical QIP as well.

In this chapter,
we start by reviewing the basic concepts of qubit and CV QIP in Sec.~\ref{sec:qubits_CVs}.
We then focus on quantum teleportation as an elementary protocol in qubit and CV QIP,
and explain its applications to quantum computing in Sec.~\ref{sec:teleportation_and_computing}.
Finally in Sec.~\ref{sec:hybrid_QIP}, we summarize our recent accomplishment of combining both technologies to realize
``hybrid'' quantum teleportation, and describe its application to hybrid QIP
that potentially overcomes the current limitations in optical QIP.

%%% Section 2 %%%%%%%%%%%%%%%%%%%%%%%%%%%%%%%%%%%%%%%%%%%%%%%%%%%%%%%%%%%%%%%%%%%%%%%%%%%%%%%%%%%%%%

\section{Qubits and continuous variables}
\label{sec:qubits_CVs}

Here we briefly review the encoding method, basic technologies, and difficulties in qubit and CV QIP.
The comparison between these two approaches are summarized in Table.~\ref{tab:qubit_CV}.

\begin{table}
\caption{Comparison between qubit and CV QIP.}
\label{tab:qubit_CV}
\begin{tabular}{p{1.5cm}p{4.3cm}p{5.5cm}}
\svhline \noalign{\smallskip}
 & Qubit QIP & Continuous-variable QIP \\
\noalign{\smallskip} \hline \noalign{\smallskip}
Carrier & Degrees of freedom of a photon  & Quadratures of a light field  \\
Basis & Photon number basis: $\{\ket{n}\}$ & Quadrature basis: $\{\ket{x}\}$ or $\{\ket{p}\}$\\
Encoding & $\ket{\psi}=\alpha\ket{0,1}+\beta\ket{1,0}$ & $\ket{\psi}=\int_{-\infty}^{\infty}\psi(x)\ket{x}dx$ \\
Source & Photons by PDC (weak pump) & Squeezed light by PDC (strong pump)\\
Detector & Photon detector (measures $\hat{n}$) & Homodyne detector (measures $\hat{x}$ or $\hat{p}$) \\
Difficulty & Two-qubit gate (\textit{e.g.} CNOT gate) & Non-Gaussian gate (\textit{e.g.} cubic phase gate) \\
\noalign{\smallskip} \svhline
\end{tabular}
\end{table}

\subsection{Qubits}
\label{subsec:qubits}

\begin{figure}[b]
% \sidecaption
\centering
\includegraphics[scale=0.62]{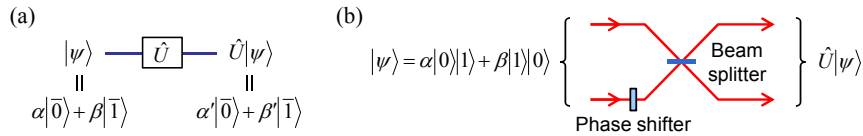}
\caption{Quantum logic gate for qubits.
(a) Circuit of a single-qubit gate $\ket{\psi}\to\hat{U}\ket{\psi}$.
(b) Implementation of a single-qubit gate for a photonic qubit encoded in two paths (spatial modes).
Appropriate choice of the phase shift and beam splitter transmissivity enables arbitrary single-qubit gates $\hat{U}$.
}
\label{fig:qubit_gate}
\end{figure}

The basic unit of information in the classical digital information processing is a bit, which can have only one of two values, `0' or `1'.
The quantum analogue of the classical bit is called a qubit, which is a superposition of the two values.
Qubit operations in optics, which exploits the particle-like discrete nature of light,
can be represented by photon number basis $\{\ket{n}\}$.
This is the eigenstate of the number operator $\hat{n}=\hat{a}^\dagger\hat{a}$ ($\hat{n}\ket{n}=n\ket{n}$ for $n=0,1,2,\ldots$),
where $\hat{a}$ and $\hat{a}^\dagger$ are annihilation and creation operators of a quantized electromagnetic field ($[\hat{a},\hat{a}^\dagger]=1$).
Usually a qubit is encoded in the degrees of freedom of a single photon (such as polarization, time of arrival or spatial modes);
it can be described using two optical modes as 
\begin{equation}
\ket{\psi}=\alpha\ket{0}\ket{1}+\beta\ket{1}\ket{0}=\alpha\ket{\bar0}+\beta\ket{\bar1}.
\end{equation}
Here $\ket{0}$ and $\ket{1}$ are vacuum (zero photon) and single photon states,
while $\ket{\bar0}$ and $\ket{\bar1}$ denote logical `0' and `1', respectively.
The information is encoded in the complex amplitudes $\alpha$ and $\beta$  ($|\alpha|^2+|\beta|^2=1$);
this is processed by sequential quantum logic gates to realize quantum computation.
A quantum logic gate can be described by the transformation $\ket{\psi}\to\hat{U}\ket{\psi}$, where $\hat{U}$ is an unitary transformation [Fig.~\ref{fig:qubit_gate}(a)].

Let us now move on to the physical implementation of qubit QIP.
Generation and measurement techniques of photonic qubits are well developed~\cite{12Pan}.
The most standard source for single photons is parametric down conversion (PDC),
where a pump photon is probabilistically converted into two photons via a nonlinear crystal.
Measurements in the logical basis can be readily implemented with photon detectors. 
The next question to follow is how to implement quantum logic gates. 
In order to realize universal qubit QIP,
arbitrary single-qubit gates and
at least one two-qubit gate are required~\cite{00Nielsen}.
The former is easily implemented with simple linear optics,
such as beam splitters and phase shifters [Fig.~\ref{fig:qubit_gate}(b)].
One example of the latter is the controlled-NOT (CNOT) gate,
which flips the state of a target qubit only if the control qubit is in the state `1'.
This is equivalent to the state of a single photon being controlled by another single photon via optical Kerr interaction
(third-order nonlinear optical effect~\cite{11vanLoock});
very large nonlinearity is required to induce this effect on a single photon. 
This makes the implementation of such two-qubit gates a major difficulty in qubit QIP.

\subsection{Continuous variables}
\label{subsec:CVs}

\begin{figure}[b]
% \sidecaption
\centering
\includegraphics[scale=0.62]{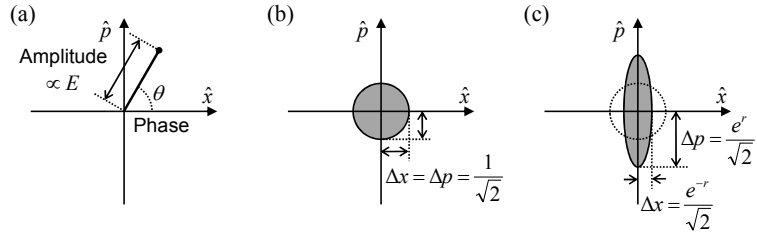}
\caption{Phase space description.
(a) Quadratures $\hat{x}$ and $\hat{p}$ correspond to $E\cos\theta$ and $E\sin\theta$ [Eq.~(\ref{eq:E_x_p})],
and these relations can be illustrated in the phase space spanned by $\hat{x}$ and $\hat{p}$.
(b) Quadrature distribution of a vacuum state.
(c) Quadrature distribution of a squeezed state.
The degree of squeezing is often characterized by the squeezing parameter $r$.
}
\label{fig:phase_space}
\end{figure}

The alternative way to encode quantum information is to use continuous basis. %instead of the discrete photon number basis.
This idea is similar to classical analog information processing,
such as AM/FM radios, where continuous values are encoded in amplitude and phase modulations of radio waves.
In CV QIP, quadratures $\hat{x}$ and $\hat{p}$ of optical waves~\cite{05Braunstein}
\begin{equation}
\hat{x}=(\hat{a}^\dagger+\hat{a})/\sqrt2,\quad
\hat{p}=i(\hat{a}^\dagger-\hat{a})/\sqrt2\quad (\hbar=1)
\end{equation}
are used to encode the superposition of continuous values. 
An intuitive definition of quadrature values would be the sine and cosine components of an oscillating wave [Fig.~\ref{fig:phase_space}(a)]:
\begin{equation}
E\sin(\omega t+\theta)=E\cos\theta\sin\omega t+E\sin\theta\cos\omega t
\propto x \sin\omega t + p \cos\omega t.
\label{eq:E_x_p}
\end{equation}
The commutation relation $[\hat{x},\hat{p}]=i$ can be derived from $[\hat{a},\hat{a}^\dagger]=1$.
Therefore all quantum states satisfy the uncertainty relation $\Delta x\Delta p\ge 1/2$.
Even the vacuum state has a so-called zero-point fluctuation of $\Delta x=\Delta p= 1/\sqrt2$,
though its quadratures are zero on average $\braket{\hat{x}}=\braket{\hat{p}}=0$ [Fig.~\ref{fig:phase_space}(b)].
The eigenstates of $\hat{x}$ and $\hat{p}$ form continuous bases $\{\ket{x}\}$ and $\{\ket{p}\}$ ($\hat{x}\ket{x}=x\ket{x}$, $\hat{p}\ket{p}=p\ket{p}$ for $x,p\in\mathbb{R}$).
An example of CV quantum information described in the $\hat{x}$-quadrature basis reads
\begin{equation}
\ket{\psi}=\int_{-\infty}^{\infty}\psi(x)\ket{x}dx.
\label{eq:CV_state}
\end{equation}
Here the information is represented by the wave 
function $\psi(x)=\braket{x|\psi}$;
which is to be processed by CV quantum logic gates $\ket{\psi}\to\hat{U}\ket{\psi}$.
Note that the state in Eq.~(\ref{eq:CV_state}) can also be expanded in the photon number basis as
$\ket{\psi}=\sum_{n=0}^\infty c_n\ket{n}$ with $c_n=\braket{n|\psi}$.
Therefore, CV QIP includes qubit QIP as a special case when the infinite dimensional Hilbert space is limited to a smaller subspace.

The key resource in implementing CV QIP is squeezed light.
The quantum noise level of one of the quadratures ($\Delta x$ or $\Delta p$) of squeezed light is below the vacuum noise level,
with infinitely squeezed light $\Delta x\to0$ ($\Delta p\to0$) corresponding to the quadrature eigenstate $\ket{x=0}$ ($\ket{p=0}$) [Fig.~\ref{fig:phase_space}(c)].
The squeezed light can be deterministically generated using the same mechanism as PDC but with a strong pump beam.
Furthermore, measurement of $\hat{x}$ and $\hat{p}$ values can be carried out with high efficiency by homodyne detectors.
Now let us move on to how to implement universal CV QIP.
In order to construct an arbitrary unitary transformation $\hat{U}=\exp(-i\hat{H}t)$,
Hamiltonians $\hat{H}$ of arbitrary polynomials of $\hat{x}$ and $\hat{p}$ are required~\cite{99Lloyd}.
Unitary transformations which involves Hamiltonians of linear or quadratic in $\hat{x}$ and $\hat{p}$
are called ``Gaussian'' gates,
which can be readily implemented by standard techniques such as beam splitters, phase shifters, squeezing and modulation.
However, CV universality requires at least one ``non-Gaussian'' gate which involves a higher order Hamiltonian,
such as the cubic phage gate $\hat{U}=\exp(i\chi \hat{x}^3)$ ($\hat{H}\propto\hat{x}^3$)~\cite{99Lloyd}.
Implementation of non-Gaussian gates is a major problem in CV QIP
as they require at least third-order optical nonlinearity;
this is hard to implement for arbitrary quantum states of light.
In this sense, non-Gaussian gates share the same difficulty as the CNOT gate in DV QIP.

%%% Section 3 %%%%%%%%%%%%%%%%%%%%%%%%%%%%%%%%%%%%%%%%%%%%%%%%%%%%%%%%%%%%%%%%%%%%%%%%%%%%%%%%%%%%%%

\section{Quantum teleportation and quantum computing}
\label{sec:teleportation_and_computing}

In optical QIP, ``quantum teleportation'', 
the transfer protocol of quantum information, plays the central role in building quantum logic gates.
This section discusses the basics and applications of quantum teleportation.

\subsection{Quantum teleportation}
\label{subsec:teleportation}

Quantum teleportation~\cite{93Bennett} is the act of transferring quantum information to distant places
without direct transmission of the physical entity itself.
Its basic concepts and implementations are as follows.

\subsubsection{Basic concept}

\begin{figure}[t]
% \sidecaption
\centering
\includegraphics[scale=0.61]{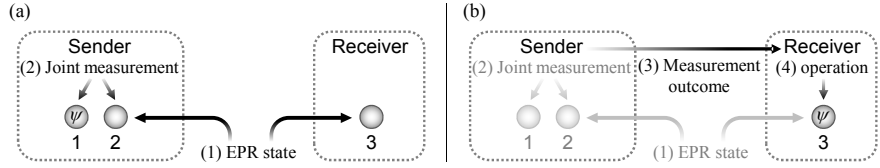}
\caption{Procedure of quantum teleportation.
An unknown quantum state $\ket{\psi}$, originally possessed by mode 1,
is teleported to mode 3 after the four steps (1-4) illustrated in the figure.}
\label{fig:teleportation_procedure}
\end{figure}

It is impossible to transfer unknown quantum superposition states from a sender to a spatially distant receiver
only via classical communications (\textit{e.g.} phone and e-mail)~\cite{93Bennett}.
However, this can be accomplished by following the quantum teleportation protocol which utilizes quantum entanglement shared between the two parties.
This idea was first proposed for qubits in 1993 by Bennett \textit{et al.}~\cite{93Bennett},
and later extended to CVs by Vaidman~\cite{94Vaidman}.
The basic procedure of quantum teleportation is the same for both schemes.
Here we define the mode of the quantum state $\ket{\psi}$ to be teleported as mode 1.
As shown in Fig.~\ref{fig:teleportation_procedure}, quantum teleportation consists of the following four steps:
\begin{enumerate}
\item[(1)] The sender and receiver share an ancillary entangled state in modes 2 and 3 (Einstein-Podolsky-Rosen state, EPR state).
\item[(2)] The sender performs a joint measurement on modes 1 and 2 (Bell-state measurement).
\item[(3)] The sender sends the measurement outcome to the receiver via classical communications.
\item[(4)] The receiver performs an unitary operation on mode 3 based on the measurement outcome;
 as a result $\ket{\psi}$ appears in mode 3.
\end{enumerate}
In this way, the quantum state $\ket{\psi}$ is transferred from mode 1 to 3 by means of the shared entanglement and classical communications.
The nomenclature of ``teleportation'' comes from the fact that the initial quantum state in mode 1 inevitably vanishes, and the same quantum state reappears in mode 3. 
In this way, quantum teleportation evades violating the no-cloning theorem, which prohibits making an exact copy of a quantum state.

\subsubsection{Qubit teleportation}

\begin{figure}[t]
% \sidecaption
\centering
\includegraphics[scale=0.62]{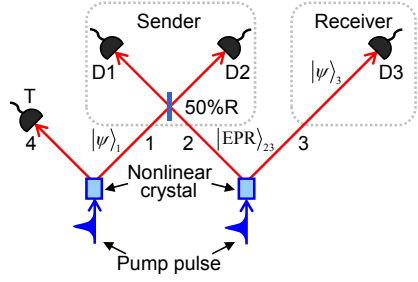}
\caption{Schematic of quantum teleportation of photonic qubits demonstrated in Ref.~\cite{97Bouwmeester}.
A polarization qubit $\ket{\psi}$ in mode 1 is prepared when T detects a photon,
and then teleported to mode 3 on condition that all of D1, D2 and D3 each detect a photon.
}
\label{fig:qubit_teleportation}
\end{figure}

After the original proposal, Bouwmeester \textit{et al.} reported the first experimental realization of quantum teleportation using photonic qubits in 1997~\cite{97Bouwmeester}.
This experiment used the polarization modes of photon 1 in Fig.~\ref{fig:qubit_teleportation} to encode the qubit:
\begin{equation}
\ket{\psi}_1=\alpha\ket{1}_{1H}\ket{0}_{1V}+\beta\ket{0}_{1H}\ket{1}_{1V}=\alpha\ket{H}_1+\beta\ket{V}_1,
\label{eq:input_qubit}
\end{equation}
where $\ket{H}_1$ and $\ket{V}_1$ denote the horizontal and vertical polarization of the photon respectively.
In this case, the ancillary EPR state in step (1) is polarization-entangled photons 2 and 3, written as
\begin{equation}
\ket{{\rm EPR}}_{23}=(\ket{H}_2\ket{V}_3-\ket{V}_2\ket{H}_3)/\sqrt2.
\end{equation}
These two photons have the following correlation: 
when one photon has horizontal (vertical) polarization, the other photon has vertical (horizontal) polarization.
Both the input qubit $\ket{\psi}_1$ and the EPR state $\ket{{\rm EPR}}_{23}$ are probabilistically created by PDC with a weak pump pulse, as shown in Fig.~\ref{fig:qubit_teleportation}.
The photon detector T is used to verify whether the input qubit had been prepared properly.
Bell-state measurement in step (2) is then performed using a 50:50 beam splitter and two photon detectors D1 and D2. 
When the two detectors simultaneously detect photons,
photons 1 and 2 are projected onto the state $\ket{\Psi}_{12}=(\ket{H}_1\ket{V}_2-\ket{V}_1\ket{H}_2)/\sqrt2$.
In this case the final state of photon 3 would read
\begin{equation}
{}_{12}\braket{\Psi|\psi}_1\ket{{\rm EPR}}_{23}=-\left(\alpha\ket{H}_3+\beta\ket{V}_3\right),
\end{equation}
which turns out to be the same polarization qubit as in Eq.~(\ref{eq:input_qubit}),
and the teleportation is completed without the operation step (4).

However, this scheme withholds two important drawbacks in terms of applications.
One is its low transfer efficiency due to the probabilistic nature of the PDC and Bell-state measurement.
The success probability is estimated to be far below 1{\%}, which does not meet the requirements for practical applications.
Another is that this scheme requires post-selection of successful events by confirming the existence of the output qubit with detector D3~\cite{98Braunstein}.
This removes the unwanted events when there is no output photon
(this event corresponds to the case when
two photon pairs are created in the left nonlinear crystal of Fig.~\ref{fig:qubit_teleportation}
and no photons in the right).
The transferred qubits are destroyed in this process, and thus cannot be used for further information processing.
Despite these inefficiencies, the transfer fidelity of the post-selected successful events are high
with the potential to reach 100\% in principle.

\subsubsection{CV teleportation}

\begin{figure}[b]
% \sidecaption
\centering
\includegraphics[scale=0.62]{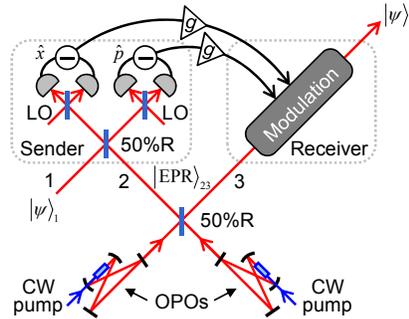}
\caption{Schematic of quantum teleportation of CVs demonstrated in Ref.~\cite{98Furusawa}.
A quantum state $\ket{\psi}$ of a light beam is deterministically teleported from mode 1 to 3
by on-demand EPR beams and complete Bell-state measurement followed by modulation.
CW, continuous wave; OPO, optical parametric oscillator; LO, local oscillator; $g$, classical channel gain.
In the standard protocol as in Ref.~\cite{98Furusawa}, the gain is set to unity.
However, it is shown that gain tuning is quite effective for the hybrid approaches (see Sec.~\ref{subsec:hybrid_teleportation}).
}
\label{fig:CV_teleportation}
\end{figure}

In 1998, Furusawa \textit{et al.} demonstrated teleportation of the quadratures of a light beam~\cite{98Furusawa}, 
following the proposal by Braunstein and Kimble~\cite{98Braunstein2}.
Here the input quantum state to be teleported is encoded in beam 1 as
$\ket{\psi}_1=\int_{-\infty}^{\infty}\psi(x)\ket{x}_1dx$ [Fig.~\ref{fig:CV_teleportation}] .
Teleportation of such states require the following 
ancillary EPR beam 2 and 3, entangled in quadrature basis:
\begin{equation}
\ket{{\rm EPR}}_{23}\propto\int_{-\infty}^{\infty}\ket{x}_2\ket{x}_3dx=\int_{-\infty}^{\infty}\ket{p}_2\ket{-p}_3dp.
\label{eq:CV_EPR}
\end{equation}
The quadratures of each of the EPR beam are quite noisy,
but these two beams behave in a correlated way: 
when beam 2 has a $\hat{x}$-quadrature value of $x$ ($\hat{p}$-quadrature value of $p$), beam 3 has the value of $x$ ($-p$).
The strength of CV teleportation is that
approximated EPR beams can be prepared on-demand by mixing two orthogonally squeezed beams
(approximated states of $\ket{x=0}$ and $\ket{p=0}$) on a 50:50 beam splitter.
These squeezed beams are deterministically generated using an optical parametric oscillator (OPO),
 a cavity-enhanced version of the PDC pumped by a strong continuous-wave beam.
Furthermore, CV Bell-state measurement can be performed completely by
two homodyne detectors that each measure either $\hat{x}$ or $\hat{p}$.
These measurements are followed by amplitude and phase modulations for step (4) to displace (shift) the quadratures of beam 3 in the phase space
according to the measured values of $\hat{x}$ and $\hat{p}$.
Intuitively, this measurement-and-modulation process cancels out the correlated quadrature noise between beams 2 and 3 in Eq.~(\ref{eq:CV_EPR}).
If the quadratures are perfectly correlated, the noise is completely canceled out,
and beam 3 becomes the same quantum state as the input state $\ket{\psi}_1$.
Since all these steps can be performed in a deterministic fashion,
a CV teleportation device can always teleport the input state, and outputs the corresponding state in beam 3.
This deterministic nature is a clear advantage over the probabilistic scheme of qubit teleportation
(see Refs.~\cite{05Braunstein,98Braunstein2} for more detailed mathematical description of CV teleportation).

The major drawback of CV teleportation is that
the transfer fidelity is limited due to the imperfect EPR beams generated from finitely-squeezed light.
More specifically, the output state is always degraded by excess noise contamination due to the imperfect quadrature correlation between beams 2 and 3.
The fidelity approaches unity in the limit of infinite squeezing, which would require infinite energy.
Though efforts were made to circumvent this drawback using higher squeezing levels, transfer errors were not  eradicated.

\subsection{Quantum computing based on quantum teleportation}
\label{subsec:tele_based_computing}

Quantum teleportation was originally proposed for transferring quantum information as it is ($\ket{\psi}\to\ket{\psi}$),
but later works have revealed a more auspicious potential:
quantum teleportation can work a quantum logic gates ($\ket{\psi}\to\hat{U}\ket{\psi}$) only with slight modification.
Below we deal with two main schemes to realize logic gates based on the CV teleportation circuit.
A similar discussion can be made for the qubit teleportation circuit.

\begin{figure}[b]
% \sidecaption
\centering
\includegraphics[scale=0.62]{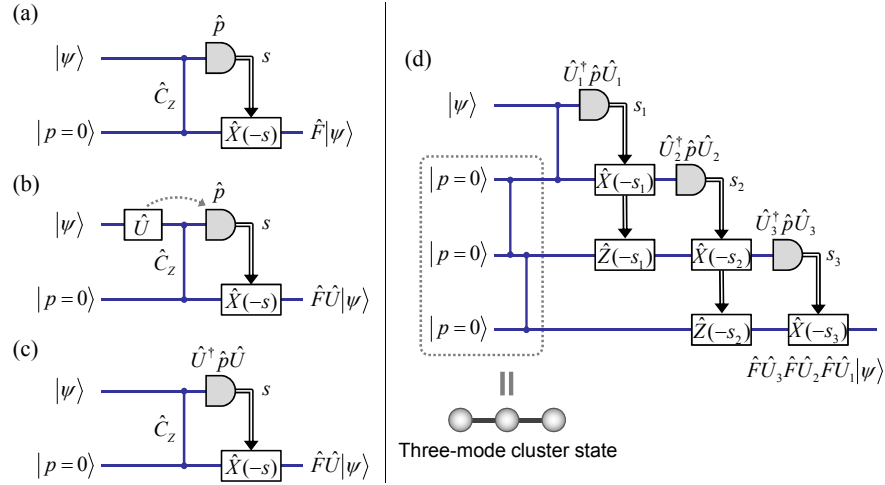}
\caption{Cluster-state quantum computation.
$\hat{C}_Z=\exp(i\hat{x}_1\hat{x}_2)$, controlled-phase gate;
$\hat{F}=\exp\left[i\pi(\hat{x}^2+\hat{p}^2)/4\right]$, Fourier transform;
$\hat{U}=\exp\left[if(\hat{x})\right]$, a desired unitary transformation;
$\hat{X}(s)=\exp(-is\hat{p})$, $\hat{x}$-displacement operation;
and $\hat{Z}(s)=\exp(is\hat{x})$, $\hat{p}$-displacement operation.
(a) An elementary CV teleportation circuit.
(b) $\hat{U}$ is applied to the input state to obtain the output $\hat{F}\hat{U}\ket{\psi}$.
(c) Measured variable is changed from $\hat{p}$ to $\hat{U}^\dagger\hat{p}\hat{U}$ to obtain the output $\hat{F}\hat{U}\ket{\psi}$.
(d) Quantum computation using a three-mode cluster state.
}
\label{fig:cluster_computation}
\end{figure}

\subsubsection{Cluster-state quantum computation}

Let us first consider an elementary CV teleportation circuit shown in Fig.~\ref{fig:cluster_computation}(a)~\cite{06Menicucci}.
The original CV teleportation in Fig.~\ref{fig:CV_teleportation} is equivalent to the case when this elementary circuit is cascaded twice.
In this circuit, an arbitrary input state $\ket{\psi}$ is first coupled with an ancillary state $\ket{p=0}$
via a controlled-phase gate $\hat{C}_Z$,
and then its $\hat{p}$ quadrature is measured by a homodyne detector.
The measurement outcome $s$ decides the amount of the $\hat{x}$-displacement operation $\hat{X}(-s)$ on the other mode.
After this operation, the Fourier transformed input state $\hat{F} \, \ket{\psi}$ appears.

Suppose an unitary operation $\hat{U}=\exp\left[if(\hat{x})\right]$ is applied to the input state
before teleportation as shown in Fig.~\ref{fig:cluster_computation}(b).
In this case, the input of the teleportation circuit is replaced by $\hat{U}\ket{\psi}$,
and therefore the output should be $\hat{F}\hat{U}\ket{\psi}$.
Since $\hat{U}$ and $\hat{C}_Z$ commute, $\hat{U}$ may be performed after the $\hat{C}_Z$ gate.
Furthermore, $\hat{U}$ can be incorporated into the measurement part by changing the measurement from $\hat{p}$ to $\hat{U}^\dagger\hat{p}\hat{U}$.
In this way, Fig.~\ref{fig:cluster_computation}(b) can be transformed into Fig.~\ref{fig:cluster_computation}(c).
This shows that an arbitrary unitary operation $\hat{U}$ can be applied to an input state
only by appropriately changing the measurement basis of the elementary teleportation circuit.
By cascading this circuit as in Fig.~\ref{fig:cluster_computation}(d), we can perform unitary operations sequentially
to obtain the desired output state $\hat{F}\hat{U}_3\hat{F}\hat{U}_2\hat{F}\hat{U}_1\ket{\psi}$.
This process can be understood as follows.
A three-mode entangled state is prepared in advance (surrounded by a gray dashed line),
and coupled to the input state $\ket{\psi}$ by $\hat{C}_Z$ gate.
Then quantum computation is performed only by appropriate choice of the measurement.
The initial multi-mode entangled state is called a cluster state.
This cluster-state quantum computation is totally different from the conventional model for quantum computation.
The conventional model requires preparation of each quantum circuit for every unitary operation,
and therefore requires different optical circuits (hardware) for different quantum computations.
In contrast, in the cluster model, the required circuit for preparing cluster states (hardware) is always the same,
but different computations can be realized by simply choosing a different measurement basis (different software).
This software-based quantum computer is the quantum analogue of the current general-purpose computer.

\begin{figure}[b]
% \sidecaption
\centering
\includegraphics[scale=0.62]{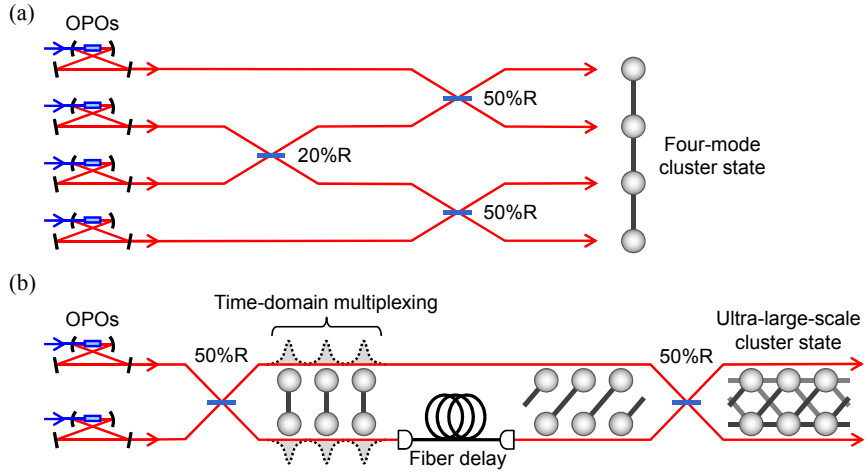}
\caption{Generation scheme of CV cluster states.
(a) A four-mode cluster state in four optical beams generated from four squeezed beams~\cite{08Yukawa}.
(b) An ultra-large-scale cluster state multiplexed in the time domain generated from two squeezed beams~\cite{13Yokoyama}.
}
\label{fig:cluster_experiment}
\end{figure}

Cluster-state quantum computation was originally proposed for qubit QIP by Raussendorf and Briegel in 2001~\cite{01Raussendorf}.
To date, preparation of few-qubit cluster states and cluster-based quantum logic gates for qubits have been reported in several experiments~\cite{05Walther,07Prevedel,08Tokunaga}.
However, due to the probabilistic nature of PDC,
preparation of large-scale cluster states are too demanding.
In contrast, CV cluster states can be generated deterministically by a scheme proposed by van Loock \textit{et al.} in 2007~\cite{07vanLoock},
which requires only mixing squeezed beams (approximated states of $\ket{p=0}$) at beam splitters with appropriate transmissivities and phases.
Figure~\ref{fig:cluster_experiment}(a) shows the schematic of generating a four-mode CV cluster state in four optical beams, demonstrated by Yukawa \textit{et al.} in 2008~\cite{08Yukawa}.
This cluster state was later used for demonstrating cluster-based one- and two-mode Gaussian gates by Ukai \textit{et al.} in 2011~\cite{11Ukai,11Ukai2}.
Though CV cluster-state computation is deterministic,
the configuration of Fig.~\ref{fig:cluster_experiment}(a) still lacks scalability
as each additional mode to the cluster state requires more OPOs and beam splitters.
In 2013, Yokoyama \textit{et al.} took a different approach to generate an ultra-large-scale CV cluster state containing more than 10,000 modes~\cite{13Yokoyama}.
Here, the modes entangled are wave packets of light in two beams, multiplexed in the time domain [Fig.~\ref{fig:cluster_experiment}(b)].
These experimental achievements show that the CV cluster state is a promising platform for CV QIP.
However, it should be noted that errors accumulate during CV cluster-state computation,
because experimentally generated cluster states are generated from finitely-squeezed states.
In addition, non-Gaussian gates required for universal CV QIP, cannot be achieved using only homodyne measurement
(solutions to this problem will be mentioned in Sec.~\ref{subsec:hybrid_computing}).

\subsubsection{Quantum gate based on off-line scheme}
\label{subsec:off_line_gate}

\begin{figure}[b]
%\sidecaption
\centering
\includegraphics[scale=0.62]{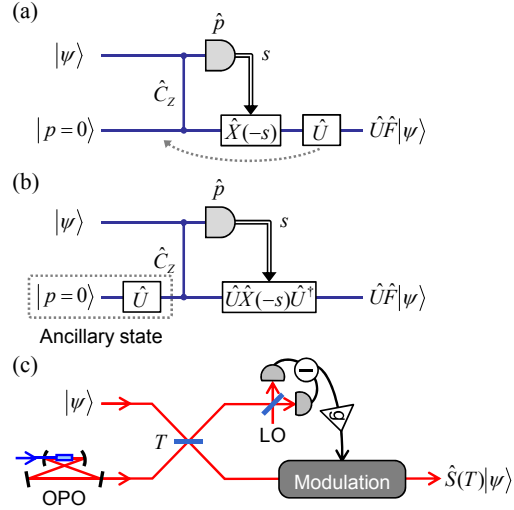}
\caption{Quantum gates based on off-line scheme.
(a) $\hat{U}$ is applied to the output state to obtain $\hat{U}\hat{F}\ket{\psi}$.
(b) Ancillary state $\hat{U}\ket{p=0}$ is used to realize the desired gate $\ket{\psi}\to\hat{U}\hat{F}\ket{\psi}$.
(c) Schematic of the universal squeezer demonstrated in Ref.~\cite{07Yoshikawa}.
Beam splitter transmissivity $T$ and gain $g$ are chosen to
perform desired degree of squeezing operation $\hat{S}(T)$ to the input state $\ket{\psi}$.}
\label{fig:gate_teleportation}
\end{figure}

Another important application of the elementary teleportation circuit are quantum gates based on off-line prepared ancillary states.
In this scheme, quantum teleportation allows fault-tolerant implementation of
difficult quantum gates that would otherwise corrupt fragile quantum information~\cite{00Nielsen,99Gottesman,03Bartlett}.
This idea dates back to the proposal of so-called ``gate teleportation'',
which was originally introduced for qubits by Gottesman and Chuang in 1999~\cite{99Gottesman},
and then extended to CVs by Bartlett and Munro in 2003~\cite{03Bartlett}.
Let us explain the basic idea by starting from the circuit of Fig.~\ref{fig:cluster_computation}(a) again.
This circuit is first extended to Fig.~\ref{fig:gate_teleportation}(a),
where the unitary operation $\hat{U}=\exp\left[if(\hat{x})\right]$ is added to the final step.
The output state in this case is $\hat{U}\hat{F}\ket{\psi}$.
By replacing $\hat{X}(-s)$ by $\hat{U}\hat{X}(-s)\hat{U}^\dagger$
and using the commutation of $\hat{U}$ and $\hat{C}_Z$,
we can move $\hat{U}$ prior to the $\hat{C}_Z$ gate, as in Fig.~\ref{fig:gate_teleportation}(b).

Importantly, when $\hat{U}$ involves a Hamiltonian of a $n$-th order polynomial of $\hat{x}$,
$\hat{U}\hat{X}(-s)\hat{U}^\dagger$ is shown to involve a Hamiltonian of $(n-1)$-th order~\cite{03Bartlett}.
In the case of $n\le3$, $\hat{U}\hat{X}(-s)\hat{U}^\dagger$ is a Gaussian gate
which is within reach of current technology.
Therefore, Fig.~\ref{fig:gate_teleportation}(b) implies the following;
once an ancillary state $\hat{U}\ket{p=0}$ is prepared,
gate $\hat{U}$ can be deterministically applied to an arbitrary input state $\ket{\psi}$
with homodyne measurement followed by a Gaussian gate,
as long as $\hat{U}$ involves a third- or lower-order Hamiltonian.
Here, the task of directly applying $\hat{U}$ to arbitrary states \textit{on-line} is replaced by
another task of preparing a specific ancillary state $\hat{U}\ket{p=0}$ \textit{off-line} prior to the actual gate,
which is much easier in experimental implementation.
In this case, gate $\hat{U}$ for $\ket{p=0}$ may be implemented in a probabilistic fashion for multiple trials until it succeeds. 
Then, only the successfully prepared ancillary states $\hat{U}\ket{p=0}$ are
stored in optical memories and consumed on demand as a resource for the logic gate
(see Sec.~\ref{subsec:hybrid_computing} for a scheme to prepare ancillary states on demand).
Note that, in contrast to the cluster-state computation where only Gaussian displacement gates are required for arbitrary $\hat{U}$ [Fig.~\ref{fig:cluster_computation}(c)],
the off-line scheme requires the gate $\hat{U}\hat{X}(-s)\hat{U}^\dagger$
and the difficulty of implementing such gate depends on $\hat{U}$.

One important example of the CV off-line scheme is the universal squeezer, a Gaussian gate 
which deterministically performs a squeezing gate to arbitrary input states $\ket{\psi}$ by means of an off-line prepared squeezed state [Fig.~\ref{fig:gate_teleportation}(c)].
The universal squeezer and quantum non-demolition (QND) sum gate based on the squeezers were already demonstrated
by Yoshikawa \textit{et al.} in 2007 and 2008, respectively~\cite{07Yoshikawa,08Yoshikawa}.
Importantly, the cubic phase gate $\hat{U}=\exp(i\chi\hat{x}^3)$ (third order, non-Gaussian)
can be implemented only with Gaussian gates
if a nonlinear cubic phase state $\exp(i\chi\hat{x}^3)\ket{p=0}$ can be prepared off-line~\cite{01Gottesman,11Marek}.
However, the experimental realization of the cubic phase gate has not yet been reported (progress towards its realization will be mentioned in Sec.~\ref{subsec:hybrid_computing}).
In the case of qubit QIP,
Gottesman and Chuang showed that the CNOT gate can also be implemented using the qubit teleportation circuit and off-line prepared ancillary states~\cite{99Gottesman}.
Linear optics quantum computing proposed by Knill, Laflamme and Milburn (KLM) in 2001~\cite{01Knill} also uses
the teleportation circuit and special ancillary states to perform two-qubit gates with near-unit success probability.
A probabilistic CNOT gate based on the KLM scheme was demonstrated by Okamoto \textit{et al.} in 2011~\cite{11Okamoto}.
Despite these proposals and demonstrations, the requirement for scalable qubit QIP will continue to be demanding, 
as long as it is based on the probabilistic generation and detection of photonic qubits.

%%% Section 4 %%%%%%%%%%%%%%%%%%%%%%%%%%%%%%%%%%%%%%%%%%%%%%%%%%%%%%%%%%%%%%%%%%%%%%%%%%%%%%%%%%%%%%

\section{Towards hybrid quantum information processing}
\label{sec:hybrid_QIP}

As mentioned above, both qubit and CV QIP come with technical problems.
The problem of qubit QIP is the low success rate,
while CV QIP has limited fidelity due to finite squeezing.
Here we introduce the recent research progress and future possibilities of
``hybrid'' QIP~\cite{11Furusawa}, which has the potential to overcome current limitations.

\subsection{Hybrid quantum teleportation}
\label{subsec:hybrid_teleportation}

Considering the fact that quantum teleportation now plays a central role in qubit and CV QIP,
quantum teleportation using a hybrid technique should be an important first step towards more advanced hybrid protocols.

\begin{figure}[b]
% \sidecaption
\centering
\includegraphics[scale=0.62]{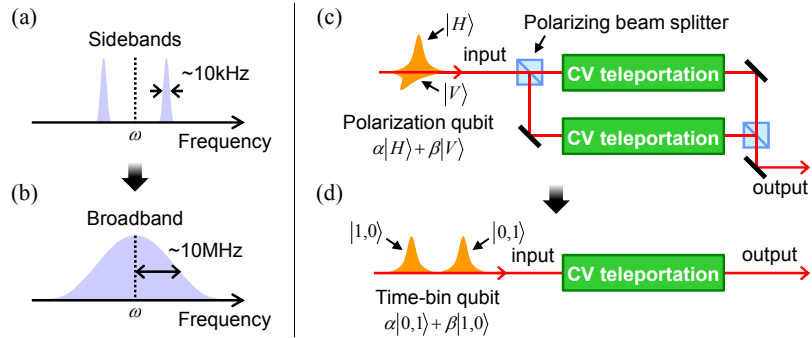}
\caption{Technologies towards hybrid quantum teleportation.
(a) Conventional CV teleportation device works on only narrow frequency sidebands around laser carrier frequency $\omega$.
(b) Broadband CV teleportation device works on frequency band with up to around 10 MHz of half-width at half maximum.
(c) CV teleportation of polarization qubits requires two teleportation devices.
(d) CV teleportation of time-bin qubits requires only one teleportation device.
}
\label{fig:hybrid_QT_development}
\end{figure}

\subsubsection{Proposal and difficulties}

One promising solution to the inefficiency of the conventional qubit teleportation scheme
is to teleport photonic qubits via a CV teleportation device.
This hybrid setting enables deterministic teleportation of qubits
by exploiting the on-demand squeezing resources and complete Bell-state measurements in the quadrature bases.
In principle, CV teleportation can be straightforwardly applied to any optical quantum state, let alone photonic qubits. 
However, experimental realization of the hybrid teleportation
was too demanding when the proposal was made in around 2000~\cite{99Polkinghorne,01Ide}.

There were three main obstacles to the experimental realization.
First was the high squeezing level requirements for the resource EPR states; these highly non-classical states were beyond the technology of that time.
Squeezing is typically quantified by the reduction in noise level of the squeezed quadrature below the shot noise level.
The world record for squeezing had been 6 dB~\cite{92Polzik}, which was not enough for such teleportation.
Takeno \textit{et al.} overcame this limitation by turning to a new nonlinear crystal, periodically poled KTiOPO$_4$; this produced 9 dB of squeezing in 2007~\cite{07Takeno}. 
The current world record for high-level-squeezing is 13 dB, reported with the same nonlinear medium by Eberle \textit{et al.}~\cite{10Eberle}.

Second was the bandwidth incompatibility. 
The typical photonic qubit has a broad bandwidth in frequency domain because it is a wave packet, i.e., a pulse.
In contrast, the conventional CV teleportation device only worked for narrow frequency sidebands~\cite{98Furusawa} [Fig.~\ref{fig:hybrid_QT_development}(a)].
Therefore it was impossible to teleport a wave packet by using the conventional setup of CV teleportation.
In order to break through such difficulty, the bandwidth of CV teleporter had to be broadened. 
Takei \textit{et al.} first broadened the bandwidth of the EPR resource in 2006~\cite{06Takei},
and then Lee \textit{et al.} used the broadband and highly entangled EPR resource to teleport highly non-classical wave packets of light in 2011~\cite{11Lee} [Fig.~\ref{fig:hybrid_QT_development}(b)].

Third, a narrow-band qubit compatible with the CV teleporter was needed.
Although the original proposals for CV teleportation of qubits
were for polarization qubits~\cite{99Polkinghorne,01Ide},
time-bin qubits were later found to be more technically compatible.
This qubit consists of two optical pulses separated temporally, and described as a superposition of a photon in
either pulse $\ket{\psi}=\alpha\ket{0,1}+\beta\ket{1,0}$ [Fig.~\ref{fig:hybrid_QT_development}(d)].
The advantage of time-bin qubits is that they can be teleported using one CV teleporter,
since the two pulses have the same polarization; polarization qubit teleportation requires two CV teleporters (one for each polarization) as in Fig.~\ref{fig:hybrid_QT_development}(c).
In 2013, Takeda \textit{et al.} developed a generation and characterization technique for time-bin qubits with a compatible frequency spectrum,
thereby completing the last piece of the hybrid teleportation system~\cite{13Takeda1}.
Now it is time for the hybrid teleportation.

\subsubsection{Demonstration of hybrid teleportation}

\begin{figure}[b]
% \sidecaption
\centering
\includegraphics[scale=0.62]{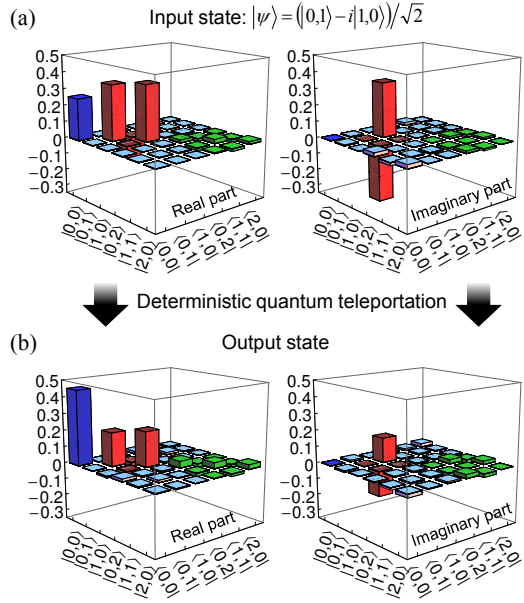}
\caption{
Experimental results of hybrid quantum teleportation in Ref.~\cite{13Takeda2}.
CV quantum teleportation is performed for qubit $\ket{\psi}=\left(\ket{0,1}-i\ket{1,0}\right)/\sqrt2$ at
squeezing parameter $r=1.01$ and gain $g\approx \tanh r$.
The two-mode density matrices are reconstructed both for the input and the output qubit states in the photon-number basis:
$\hat{\rho}=\sum_{k,l,m,n=0}^\infty\rho_{klmn}\ket{k,l}\bra{m,n}$.
}
\label{fig:hybrid_QT_result}
\end{figure}

The combination of all these state-of-the-art technologies enabled CV quantum teleportation of time-bin qubits.
Using this hybrid setup, Takeda \textit{et al.} demonstrated, for the first time,
deterministic quantum teleportation of photonic qubits in 2013~\cite{13Takeda2}.
This experiment demonstrated that, even with finite squeezing resources,
qubit information can be teleported faithfully by adjusting the classical channel gain in CV teleportation ($g$ in Fig.~\ref{fig:CV_teleportation}).
The mechanism is as follows.
For finite squeezing parameter $r$ [defined in Fig.~\ref{fig:phase_space}(c)], the standard CV teleportation protocol with $g=1$
yields a largely distorted output qubit with additional photons in general.
In contrast, a CV teleporter with $g=\tanh r$ becomes equivalent to a pure loss channel,
which only adds extra loss of $(1-\tanh^2r)$ to the input state~\cite{99Polkinghorne}.
Moreover, the single-photon-based qubit $\ket{\psi}=\alpha\ket{0,1}+\beta\ket{1,0}$
represents a quantum error detection code against photon loss,
where either a photon-loss error occurs, erasing the qubit, or a symmetric
amplitude damping leaves the input qubit state completely intact~\cite{00Nielsen}.
These two facts together mean that the CV teleporter transforms
the initial qubit state as
\begin{equation}
\ket{\psi}\!\bra{\psi}\longrightarrow \tanh^2r\ket{\psi}\!\bra{\psi}+(1-\tanh^2r)\ket{0,0}\!\bra{0,0}.
\end{equation}
Importantly, no additional photons are created, and the qubit information $\ket{\psi}$ remains undisturbed regardless of the squeezing level.
The teleporter only adds an extra two-mode vacuum term.
Thus the weakness of CV teleportation due to the finite squeezing can be circumvented to a great extent by gain tuning.

One of the experimental results are shown in Fig.~\ref{fig:hybrid_QT_result}.
The qubit components in the subspace spanned by $\{\ket{0,1},\ket{1,0}\}$ decrease from $69\%$ at the input state to $42\%$ at the output, 
due to the extra loss added by the teleporter.
However, the output qubit components still retain the original phase information of the superposition of $\ket{0,1}$ and $\ket{1,0}$ at input, 
demonstrating that the qubit information is faithfully teleported.
The overall transfer fidelity ranged from 79 to 82\% for four different qubits, all of which exceed the classical limit of teleportation.
It was later shown that these experimental results are in good agreement with its corresponding theoretical model~\cite{13Takeda3}.
By extension of this setup, Takeda \textit{et al.} also performed
CV quantum teleportation of discrete-variable entanglement in the form of a photon split by a beam splitter;
this demonstrated the genuine quantum nature of the hybrid teleportation system~\cite{13TakedaFiO}.

\subsection{Hybrid quantum computing}
\label{subsec:hybrid_computing}

The CV teleportation circuit has now become compatible with the basic technologies of qubit QIP,
such as pulsed single photons and photon counting measurements.
The combination of deterministic gates based on CV teleportation
and nonlinear optical resources in qubit QIP
potentially gives us great benefit for implementing universal quantum computers in both CV and qubit regimes.

\subsubsection{Hybrid approach to CV universality}

\begin{figure}[t]
% \sidecaption
\centering
\includegraphics[scale=0.62]{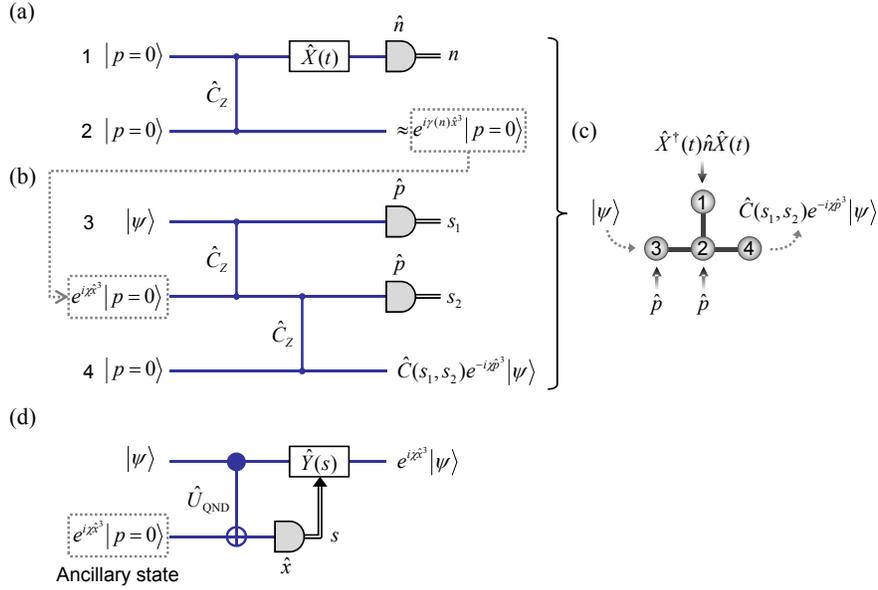}
\caption{
Implementation of a cubic phase gate.
(a) Displacement operation $\hat{X}(t)$ with sufficiently large $t$ and
photon number measurement produces an approximate version of a cubic phase state~\cite{01Gottesman}.
The cubic phase state depends on the measurement outcome $n$, and
for the desired cubic phase gate, additional squeezing operations are needed.
(b) The prepared cubic phase state is used to perform the cubic phase gate to an arbitrary input state $\ket{\psi}$.
Gaussian operations can undo the operator $\hat{C}(s_1,s_2)$ depending on the homodyne results $s_1$ and $s_2$.
(c) The cubic phase gate can be implemented based on a CV cluster state, homodyne measurement and
a nonlinear measurement onto the displaced number basis $\{\hat{X}^\dagger(t)\ket{n}\}$~\cite{09Gu}.
(b) Cubic phase gate can be performed on the input state $\ket{\psi}$
by the off-line scheme with an ancillary cubic phase state and a Gaussian gate $\hat{Y}(s)$~\cite{01Gottesman}.
}
\label{fig:cubic_phase_gate}
\end{figure}

One challenging task towards universal CV QIP is the implementation of non-Gaussian gates, such as the cubic phase gate $\hat{U}=\exp(i\chi \hat{x}^3)$.
One non-Gaussian gate, together with already well-developed Gaussian gates,
is sufficient for realizing universal CV QIP, in principle~\cite{99Lloyd}.
In order to generate the ancilla for the cubic phase gate, Gottesman, Kitaev and Preskill (GKP)
proposed an approximate version of the cubic phase state, 
generated by squeezed states and photon counting measurements as in Fig.~\ref{fig:cubic_phase_gate}(a)~\cite{01Gottesman}.
The generated cubic phase state can be used to perform the cubic phase gate to an arbitrary input state $\ket{\psi}$
through the circuit of Fig.~\ref{fig:cubic_phase_gate}(b).
These two circuits together can be interpreted as a CV cluster-state computation using homodyne measurement and photon counting measurement
as in Fig.~\ref{fig:cubic_phase_gate}(c)~\cite{09Gu}.
Therefore the hybrid technology developed thus far
may be beneficial for realizing universal CV cluster-state computation.

Another approach to the cubic phase gate is based on the off-line scheme using an ancillary cubic phase state $\exp(i\chi \hat{x}^3)\ket{p=0}$,
as already mentioned in Sec.~\ref{subsec:off_line_gate}.
Figure~\ref{fig:cubic_phase_gate}(d) shows one possible implementation proposed by GKP~\cite{01Gottesman}.
In this implementation, all the components except for the cubic phase state are already technologically available.
When the desired gate is weak ($\chi\ll1$), 
a certain superposition state of up-to three photons becomes enough for the ancillary state, as proposed by Marek \textit{et al.} in 2011\cite{11Marek}.
This type of ancillary state has already been generated experimentally by Yukawa \textit{et al.} in 2013, albeit probabilistically~\cite{13Yukawa,13Yukawa2}.
For a deterministic cubic phase gate, this ancillary state needs to be prepared on demand.
The on-demand generation technique of non-classical optical states has been reported by Yoshikawa \textit{et al.} in 2013~\cite{13Yoshikawa}.
In this experiment, single photons are created and stored inside an OPO,
and finally released on demand through a dynamical tuning of the output coupling.
This scheme can be potentially used to prepare cubic phase states on demand.
All the ingredients essential for a deterministic cubic phase gate have become available in principle, awaiting for their future ingratiation. 

\subsubsection{Hybrid approach to qubit universality}

Once the deterministic cubic phase gate is realized, in combination with other Gaussian gates, 
the CNOT gate for qubits may be implemented deterministically.
This is because the unitary transformation of optical Kerr interaction $\hat{U}=\exp(i\chi\hat{a}^\dagger_1\hat{a}_1\hat{a}^\dagger_2\hat{a}_2)$,
which is the essence of the CNOT gate,
can be decomposed into the sequence of several cubic phase gates and other Gaussian gates~\cite{13Sefi}.
Therefore, hybrid technologies ultimately lead to universal qubit QIP,
where photonic qubits are processed by CV cluster-state computation or CV off-line scheme.
Such a hybrid implementation should be much more efficient and faster than the previous counterpart of qubit QIP,
which is solely based on probabilistic and post-selective resources and measurements.
Furthermore, the limitation of gate fidelity, which had been the weak point inherent in CV QIP,
may be circumvented by effective gain tuning used in the hybrid teleportation experiment.
As an example of hybrid quantum computing,
the squeezing operation (Gaussian) on single photons was demonstrated by Miwa \textit{et al.} in 2012~\cite{12Miwa} using the CV universal squeezer in Fig.~\ref{fig:gate_teleportation}(c).
Future technical developments would enable
deterministic logic gates for photonic qubits by a CV scheme.

In order to perform universal QIP fault-tolerantly,
GKP proposed to encode a logical qubit into the superposition of $\hat{x}$-eigenstates
as $\ket{\bar{j}}\propto\sum_{s=-\infty}^\infty\ket{x=(2s+j)\sqrt{\pi}}$ ($j=0,1$)~\cite{01Gottesman}.
This hybrid encoding is intended to protect a logical qubit against small errors
such as random shift in the quadrature variables $\hat{x}$ and $\hat{p}$.
A later work by Menicucci showed that
fault-tolerant quantum computation based on the GKP encoding is possible
by using finitely-squeezed resources above a threshold value of 20.5 dB and performing CV cluster-state computation with error correction~\cite{13Menicucci}.
Though the GKP encoding may still be far from implementable,
it offers unique and interesting concepts to optical QIP
and reveals a high potential of the hybrid approach.

%%% Section 5 %%%%%%%%%%%%%%%%%%%%%%%%%%%%%%%%%%%%%%%%%%%%%%%%%%%%%%%%%%%%%%%%%%%%%%%%%%%%%%%%%%%%%%

\section{Conclusion}
\label{sec:conclusion}

Until recent years, qubit and CV QIP had developed separately
each utilizing the quantum teleportation circuit as a key building block.
The gap between these two approaches had been wide due to the incompatibilities in experimental technologies.
However, these recent advances in combining both technologies have changed the situation.
Especially, the realization of hybrid quantum teleportation in 2013~\cite{13Takeda2}
must be a significant turning point in the development of optical QIP.
This work presents a prototype technology for hybrid QIP systems,
and will stimulate the further development of hybrid protocols to overcome the current limitations in optical QIP.

In our opinion, the hybrid approach to optical QIP will be the most promising one in the near future.
In principle, universal and fault-tolerant quantum computing can be attained
by using CV cluster-state computation and off-line schemes, 
while introducing nonlinearity through photon-counting measurement and nonlinear optical resources such as photon number states.
For scalable implementations of hybrid QIP,
multiplexing quantum modes in the time domain may be a key technology.
This idea has already been used for generating ultra-large-scale CV cluster states~\cite{13Yokoyama},
as well as for implementing hybrid quantum teleportation using time-bin encoding of a qubit~\cite{13Takeda2}.
In addition, hybrid QIP using 
on-chip integrated photonic circuits~\cite{08Politi} would be desirable.
Such an integrated architecture will decrease the size and complexity of the experimental setup,
leading to a low-loss, robust and scalable hybrid QIP.

\end{document}